\documentclass{aastex61}
\RequirePackage{natbib}

\usepackage{newtxtext,newtxmath}
\usepackage[T1]{fontenc}
\usepackage{ae,aecompl}
\usepackage{graphicx}	% Including figure files
\usepackage{amsmath}	% Advanced maths commands
\newcommand{\real}{\mathrm{Real}}
\newcommand{\sgn}{\mathrm{sgn}}

\received{}
\revised{}
\submitjournal{PASP}

\newenvironment{proof}[1][Proof]{\noindent\textbf{#1.} }{\ \rule{0.5em}{0.5em}}

\shorttitle{Logarithmic cumulants}
\shortauthors{Arg\"ueso et al.}

\begin{document}
\label{firstpage}

\title{Confusion noise due to clustered extragalactic point sources. \\ 
Application of logarithmic cumulants for parameter estimation}

\correspondingauthor{Francisco Arg\"ueso}
\email{argueso@uniovi.es}

  \author{Francisco Arg\"ueso}
  \affil{Departamento de Matem\'aticas, Universidad de Oviedo,  
  c/ Federico Garc\'\i a Lorca, 18, 33007, Oviedo, Spain}
  
  \author{Diego Herranz}
  \affil{Instituto de F\'\i sica de Cantabria (CSIC-UC), 
  Av. los Castros s/n, 39005, Santander, Spain}
  
  \author{Luigi Toffolatti}
   \affil{Departamento de F\'\i sica, Universidad de Oviedo, 
   c/ Federico Garc\'\i a Lorca, 18, 33007, Oviedo, Spain}
   
  \author{Joaqu\'\i n Gonz\'alez-Nuevo}
   \affil{Departamento de F\'\i sica, Universidad de Oviedo, 
   c/ Federico Garc\'\i a Lorca, 18, 33007, Oviedo, Spain}

%\maketitle

\begin{abstract} 
The calculation of the characteristic function of the signal fluctuations due to clustered astrophysical sources is performed in this paper. For the typical case of power-law differential number counts and two-point angular correlation function, we present an extension of Zolotarev's theorem that allows us to compute the cumulants of the logarithm of the absolute value of the intensity. 
As a test, simulations based on recent observations of radio galaxies are then carried out, showing that these cumulants can be very useful for determining the fundamental parameters defining the number counts and the correlation. If the angular correlation scale of the observed source population is known, the method presented here is able to obtain estimators of the amplitude and slope of the power-law number counts with mean absolute errors that are one order of magnitude better than previous techniques, that did not take into account the correlation. Even if the scale of correlation is not well known, the method is able to estimate it and still performs much better than if the effect of correlations is not considered. 
\end{abstract}


\section{Introduction}

One of the main goals of modern Astronomy is the study of the formation and evolution of galaxies, as they constitute an extremely useful probe for Cosmology --the study of the origin, structure and ultimate fate of the whole Universe-- and for extragalactic astronomy. Modern telescopes are able to reach epochs at which the Universe was ten percent of its present age and to observe billions of galaxies; these numbers make it imperative for astronomers to study galaxies statistically. 
       In absence of distance information, extragalactic point sources (i.e., galaxies seen
as point-like objects inside the observational beam, usually assumed as a circular Gaussian 
distribution whose angular dispersion determines its Full Width Half Maximum, FWHM)
appear as luminous spots in the particular sky area actually surveyed. Their observed flux density 
depends on many factors including the distance to that particular galaxy, its evolutionary
stage, the way part or all of it has been obscured by matter along the line of sight, etc.  
       As a result of these many effects, the distribution of the number of galaxies as a function 
of their observed flux densities, $S(\nu)$, (differential number counts,dN/dS) is a very useful cosmological probe. 
In many cases of interest, and at least for some ranges of flux densities, the differential number counts can be well approximated by a simple power law $dN/dS=R(S)=k\,S^{-\eta}$ where the amplitude $k$ and the slope $\eta$ contain a wealth of information about the physical and statistical properties of the underlying galaxy populations, and hence of the whole Universe.

Differential number counts of extragalactic sources are usually 
       measured (or estimated) by different analysis techniques,
       depending on the sensitivity of the detection instrument and 
       on the observational band of interest: direct resolved detection, 
       Bayesian detection, stacking analysis or
       probability of deflection analysis, $P(D)$, 
       \citep[see, e.g.][etc.]{dezotti10,Dole04,Berta11,PlanckXIII,PlanckVII,Harrison16}.
       Although the amplitude and slope of the number counts are usually known (or at last predictable according to current models of galaxy formation and evolution) for a wide range of wavelengths and at high flux densities, there are still many remaining gaps in our current knowledge, in particular at very faint flux densities, that are much more difficult to reach.
Thus, astronomers who try to determine these two numbers across different regions of the observable electromagnetic spectrum can encounter great difficulties in doing that.

To make things more interesting, due to the fundamental law of diffraction and also to inevitable instrumental limitations, telescopes cannot observe the sky with infinite image definition.  Any point of the sky observed through a telescope is blurred by the \textit{point spread function} of such telescope.
With a given beam, the final signal in a sky image is a mixing of bright and faint sources in which the brightest ones are still individually detectable above a given detection limit, $S_{lim}$. Sources at flux densities below this limit generate the so called \emph{confusion noise} \citep[see, e.g.,][]{Scheuer57}. 

The probability distribution of the measured intensity, $f(I)$, where I is the signal response at a point produced by the sum of source fluxes convolved with the beam \footnote{$I(x)=\Sigma \, S_n \, h(x-x_n)$ with $h$ the beam profile and $x_n$ the position of the source with flux $S_n$}, is also called $P(D)$ in the astrophysical jargon, and gives the total contribution of sources above and below the detection limit. However, P(D) is specially useful to estimate and partially reconstruct the number counts at very faint flux densities, below the detection limit of the instrument, where they cannot be directly measured.This situation is very common in astronomical images and it has been studied first at radio and X-ray frequencies.

The $P(D)$ distribution given by unresolved sources is strongly non-Gaussian and shows a long, positive tail. A wealth of information is encoded in the indistinct part of the distribution far from the tail. Hence the interest in Astronomy for the study and statistical characterization of this confusion noise. 

The confusion noise phenomenon has been widely studied in the astronomical literature since the pioneering work of \cite{Scheuer57}.
That work was
subsequently expanded by \cite{Hewish61}.
\cite{Condon74} studied the signal observed with a beam for differential number counts that follow a power law and are uncorrelated,
$ dN/dS=R(S)=k\,S^{-\eta}$. He wrote an analytical expression, formula (8) of his paper, for the characteristic function $\phi(\omega)$ of the observed intensity. 
However, the associated probability density function (pdf) $f(I)$, defined as the inverse Fourier transform of  $\phi(\omega)$, cannot be calculated analytically. 
Many years later, \cite{Herranz:2004} showed that Condon's characteristic function was a particular case of an alpha-stable distribution
$S_\alpha(\beta,\gamma,\mu)$, first introduced in \cite{Levy}:  
\begin{equation}
\phi(\omega)=\exp\left[-i\mu\omega-\gamma|\omega|^{\alpha}\left(1+i\beta \, sgn(\omega) \tan(\alpha \pi/2)\right)\right],
\end{equation}
\noindent
For a general alpha-stable distribution: $ \gamma>0 ,\, -1 \leq \beta \leq 1$ and $0 < \alpha \leq 2$ . In our case,  $\alpha=\eta-1$, $\beta$ is equal to one\footnote{That is, the confusion noise generated by a power law-distributed population of compact sources corresponds to the particular case of an alpha-stable distribution of maximum positive skewness parameter. For $\alpha=1$ the alpha-stable expression is different from (1)} 
and $\mu$ and $\gamma$ can be obtained in a simple way from $k$, $\eta$ and the effective beam size of the experiment 
\begin{equation}
\Omega_e=\int{h(\theta,\phi)^{\eta-1}\, d\Omega},
\end{equation}
\noindent
 where $h(\theta,\phi)$ is the beam profile. See formulas (10) and (11) in \cite{Herranz:2004}.
Taking advantage of the theory of alpha-stable distributions (\cite{Levy,Zolotarev,Shao,Kuruoglu01,Samoradnitsky17}), Herranz et al. also 
defined suitable logarithmic moment estimators based on Zolotarev's theorem (this theorem allowed the calculation of the expectation of $ |I|^p $, $ E\left(|I|^p\right)$). Then, they applied these estimators to the determination of $k$ and $\eta$ for power law number counts.

However, the presence of clustered --that is, spatially correlated-- sources, so common in Astronomy, is not taken into account in any of the cited works. \cite{Barcons92} wrote a pioneering work in which he found a general expression for the  characteristic function of the observed intensity, considering $n$-point correlations. In a similar way, by using random field theory, \cite{Takeuchi} wrote the Laplace transform of the same pdf.
The outline of the paper is as follows. In Section~\ref{sec:counts} we briefly present and discuss the differential number counts of astrophysical (extragalactic) sources and their correlation functions. Then,
inspired by the two last above-quoted papers, we manage to obtain a general $\phi(\omega)$ for the case of power law counts and general correlations in Section~\ref{sec:characteristic}. In order to obtain precise results, finally we only consider two-point correlations and prove that, in this case, the observed intensity can be written as the sum of two independent random variables that follow alpha-stable distributions with parameters $\alpha$ and $2\alpha$.
In Section~\ref{sec:logarithmic}, we generalize Zolotarev's theorem for the clustering case and calculate the corresponding logarithmic moments and cumulants. The proof of this generalization is provided in Appendix~\ref{sec:appendixA}. In Section~\ref{sec:symmetric}, we explore the symmetric case, defined by subtracting intensities at different points, distant enough to maintain statistical independence. This symmetrization enables us to simplify the formulas.  In Section~\ref{sec:simulations}, we illustrate with simulations the usefulness of the logarithmic cumulants to compute the parameters defining the counts and the correlation. Finally, we draw our main conclusions in section~\ref{sec:conclusions}.
 
 \vspace{5mm}
 
 \section{Statistics of the distribution of extragalactic point sources} \label{sec:counts}
 
 \vspace{5mm}
 
 \subsection{Differential number counts}

%\footnote{Differential number counts, $dN/dS(\nu)$
%--being $S(\nu)$ the flux density 
%and $\nu$ the observed  
%frequency band--
%are defined as the number of extragalactic sources per steradian %observed with flux densities between $S_i$ and  $S_i+dS$ per unit %flux-density interval. 
%The bin $[S_i, S_i+dS]$ is one of the $n$ flux density bins into which %the total observed interval can be divided. 
%This total interval is fixed by the specific technical characteristics %of the telescope used. Differential number counts can usually be fit 
%with a power law, 
%$dN/dS (\nu) \propto k S^{-\eta}$,
%at least in a limited flux density interval.}, 
The study of differential $dN/dS(\nu)$, and of integral, $N(>S(\nu))$, number counts of extragalactic point sources (i.e., galaxies seen as point-like objects in the sky given that their projected angular dimension is less than the beam size of the telescope/radio-telescope) is a field of great historical relevance in astrophysics. In fact, these statistics can be used to efficiently constrain the {\it evolution with cosmic time} of the observed galaxy populations at each selected frequency bands. This is easily performed by comparing the slope, $\eta$, of the {\it observed number counts} with the one foreseen for the static Euclidean universe, $\eta=2.5$, filled with galaxies uniformly distributed in the surveyed volume, which constitutes the benchmark comparison model. Any deviation in the observed number counts from the predictions of the static Euclidean universe implies evolution with cosmic time of the underlying galaxy population (see, e.g., \cite{Longair66} for a general and basic discussion on this fundamental comparison and on other related topics; see also \cite{Blain93} for an interesting application of the number counts distribution and of related statistics to the submillimeter wavebands of the electromagnetic spectrum).

   The rapid development of radio antennas after the Second World War allowed that number counts of radio selected
 --at cm to mm wavelengths--
 extragalactic sources were the first ones to be measured: the very first results can be dated back to \cite{Mills52}. Soon after, number counts of radio sources achieved instant notoriety when \cite{Ryle55} and \cite{RyleSch55} announced that the slope of the observed counts at very bright flux densities ($S(\nu)\geq 1 $ Jy) was steeper 
--i.e., $\eta > 2.5$-- than the one corresponding to a static Euclidean universe, that implied extragalactic radio sources must be evolving with cosmic time in space density or luminosity. Since then, many more surveys have been carried out to measure the source counts at various radio frequencies, both extended to the whole sky or to a limited sky area, which confirmed the importance of cosmological evolution (see, e.g., \cite{dezotti10} for a recent and comprehensive review on the subject).

More recent works at mm/sub-mm wavelengths or in the far-infrared domain of the electromagnetic spectrum \citep[see, e.g.,][]{scott12,hatsukade13b,hatsukade13,carniani15,Vernstrom,Whittam17a,Whittam17b}
show that the differential number counts of faint extragalactic point sources can be well fitted usually by sub-euclidean count slopes $\eta \leq 2.5$ down to sub-mJy or even microJansky fluxes.  
 Investigating the number counts at these faint fluxes is important for understanding the evolution of sources at low luminosities and/or high redshifts $(z > 2)$. There is an open debate about what might be happening at flux densities fainter than the current limits ($S(\nu) < 1$-10 $\mu$Jy) of the source number counts in the radio bands 
 (see the works of \cite{Condon12} and \cite{Vernstrom} at 3 GHz). New source populations, related to the epoch of galaxy formation, could be detected and their early cosmological evolution could be studied by the analysis of sub-$\mu$Jy number counts.
     Recent works in this field  will be used in Section 6 to simulate fluxes in order to test our new technique for the determination of $k$ and $\eta$. 
\vspace{10 mm}

 \subsection{n-point and 2-point angular correlation functions}

The angular distribution of sources on the celestial sphere is represented mathematically by a random field, $\rho(\mathbf{x})$, with  $\rho(\mathbf{x})$ the number of sources per solid angle unit at a given point. All the information about the field is given by its mean $\mu=<\rho(\mathbf{x}>=\frac{N}{4\pi}$, with N the total number of sources (we assume that the field is homogeneous), the two-point correlation function $\omega(\mathbf{x_1},\mathbf{x_2})$, related to the two-point moment by  

\begin{equation} \label{eq:corr1}
<\rho(\mathbf{x_1})\rho(\mathbf{x_2})>=\mu^2(1+\omega(\mathbf{x_1},\mathbf{x_2}))
\end{equation}
\noindent
and the n-point correlations $\omega_n(\mathbf{x_1},.........,\mathbf{x_n})$, that can be expressed in terms of higher-order cumulants of the field,  \citep[see][]{Barcons92}, and are zero unless all positions at which they are calculated are correlated.

Few data have been so far collected on the correlations among galaxies of order higher than two 
\citep[see, e.g.,][for the IRAS 1.2 Jy sample]{Meiksin92}.
This is obviously due to the lack of sufficient data (point sources) to calculate /estimate three- or n-point correlation functions than could be reliable, although in a limited angular scale interval. Historically, the determination of the three-point correlation function (3PCF) of galaxies was pioneered by 
\cite{Peebles75}
using the Lick and Zwicky angular catalogs of galaxies. After this first outcome, among other results we can single out \cite{Verde02} that computed the bispectrum --i.e. the Fourier counterpart of the 3PCF -- using galaxies in the 2-Degree Field Galaxy Redshift Survey, and concluded that the non-linear bias was consistent with zero.
More recent results still seem to find consistency with the usual hierarchical clustering model --the process by which larger structures are formed through the continuous merging of smaller structures-- but it will be necessary to rely   on future very large surveys (e.g., Euclid satellite, Large Synoptic Sky Telescope (LSST), ALMA, JWST, etc.) for obtaining new and more conclusive data on the 3PCF 
\citep[see, e.g.,][]{Slepian15}.

In view of the relative paucity of current data on the 3PCF and on higher order correlations, although we will write a general formula for the $n$-point correlation case, we will basically work with the two-point correlation function, that is, we neglect all higher-order contributions.
The two-point correlation only depends on the angular distance $\theta$, --i.e. the underlying field is assumed to be homogeneous and isotropic--, and has been usually approximated by a power-law with a characteristic correlation angular distance $\theta_0$ 
\begin{equation}
\omega(\theta)=\left(\frac{\theta}{\theta_0}\right)^{-\delta}
\end{equation}
\noindent
after the first evidence of it was found by \cite{Totsuji} and subsequently confirmed, with scaling relations demonstrated, by \cite{Groth77}.
For instance, \cite{loan97} found $\omega(\theta)=0.01 \theta^{-0.8}$,
for sources at 5 GHz and  
\cite{Blake2002}
obtained $\omega(\theta)=0.001\, \theta^{-0.8}$, that is $\theta_0=0.64''$, at 1.4 GHz. More recently, \cite{Vernstrom} approximated the correlation of sources at 3 GHz with a similar power-law $\omega(\theta)=(\theta/\theta_0)^{-0.8}$ but with $\theta_0=0.06''$ .

These results at radio frequencies are also confirmed in other frequency bands of the electromagnetic spectrum. For example, recent measurements of the angular correlation function of sub-millimeter galaxies identified in four of the five fields observed in the \textit{Herschel} Astrophysical Terahertz Large Area Survey works
\citep[H-ATLAS,][]{HATLAS}
show that a power-law approximation of $\omega(\theta)$ with $\delta=0.8$ is still a very good approximation up to the highest redshift interval in which a signal is detected
\citep{amvrosiadis}. In all these cases, $\theta_0$, the characteristic correlation angle, is the only parameter that changes whereas $\delta= 0.8$ remains fixed, which is related to the usual slope of the 3-dimensional spatial correlation, $\gamma = 1.8$ \citep[see, e.g.,][]{davis77}. Therefore, although our general results do not depend on the specific value of $\delta$, we will use $\delta =0.8$ in our simulations and different values of $\theta_0$, chosen from relevant observations.

\vspace{5mm}

\section{Characteristic function of the observed signal distribution} \label{sec:characteristic}

As commented before, we try to calculate the confusion noise produced by filtered point sources when 
we include the effect of clustered sources. A general approach to this problem was developed in \cite{Takeuchi} by using the mathematical techniques of point field theory. They were able to write a general expression, formula (46) of their paper, for the Laplace transform of the probability density function (pdf) of the intensity fluctuations, $f(I)$, being $I$ the signal response observed with a certain beam. 

We use a similar formula to that of \cite{Takeuchi}, but for the Fourier transform (characteristic function), since we intend to obtain a generalization for clustered sources of the results found in \cite{Herranz:2004} for the case of unclustered sources. Therefore, our formula can be written
\begin{equation} \label{eq:char1}
\phi(\omega)=\exp \left[\sum_{n=1}^{\infty}  \frac{1}{n!} \int_{\Omega_b}
\,\ldots \int_{\Omega_b} \prod_{j=1}^{n}\, \int_{S_j}(e^{-i\omega S_j h(\mathbf{x_j})}-1)\,R(S_j)\, dS_j \,w_n(\mathbf{x_1},\ldots,\mathbf{x_n})
\,\mathbf{dx_1} \ldots \mathbf{dx_n}\right],
\end{equation}
\noindent
where n is the number of points (positions of sources $\mathbf{x_1},...,\mathbf{x_n}$ ) involved in the n integrals over the beams, taking into account the n-point correlation; and j, that goes from 1 to n, indicates the flux and position of each individual source. We multiply the integrals over the fluxes, with $R(S_j)$ the differential source number counts with a flux density $S_j$, $h(\mathbf{x_j})$ the beam pattern of the instrument, $\Omega_b $ the beam area and $w_n(\mathbf{x_1},...,\mathbf{x_n})$ the $n$-point angular correlation, that encapsulates the statistical information about the spatial distribution of the sources, \citep[see][]{Barcons92}.
To summarize, we calculate the exponential of a sum of integrals that include, in principle, the contribution of all n-point correlations. This sum is the characteristic function of the intensity , $I$.

From (\ref{eq:char1}) we can find a simpler expression if we assume that the fluxes follow a power-law distribution $ R(S)=k S^{-\eta}$. When we introduce this power-law formula in (\ref{eq:char1}) and perform the change of variable $ S_j h(\mathbf{x_j})=s_j$ we have
\begin{equation} \label{eq:char2}
\phi(\omega)=\exp\left[\sum_{n=1}^{\infty}  \frac{1}{n!}\int_{\Omega_b}
\ldots\int_{\Omega_b} \prod_{j=1}^{n} h^{\eta-1}(\mathbf{x_j}) w_n(\mathbf{x_1},\ldots,\mathbf{x_n})   
\mathbf{dx_1} \ldots \mathbf{dx_n} \prod_{j=1}^{n} \int_{s_j}(e^{-i\omega s_j }-1) k s_j^{-\eta}ds_j\right].
\end{equation}
\noindent
The integral in $ s_j$ can be readily calculated
\begin{equation}
\int_{s}(e^{-i\omega s }-1)\,k s^{-\eta}\, ds=k|\omega|^{\eta-1}\Gamma(1-\eta)\left(\sin{\pi \eta/2}-i\, \sgn(\omega)\cos{\pi \eta/2}\right),
\end{equation} 
\noindent
where $\Gamma$ is the gamma function. The integral of the real part converges for $1< \eta < 3$ and the integral of the imaginary part for $0< \eta < 2$. Therefore, both integrals are well defined for $1< \eta < 2$.
In the case without correlations, if we subtract independent intensities, we obtain a symmetrized random variable and are able to remove the imaginary part. Then, the characteristic function is real and if we define $ \alpha=\eta-1$, it is an alpha-stable distribution within the standard range $0< \alpha < 2$. 
 However, in the general clustering case we have to resign ourselves to the convergence for $0< \alpha < 1$.

 %if we subtract the average intensity, the latter integral converges for $2< \eta < 4$ if n=1 (no correlations) and we obtain a simple expression. The snag is %that this is true for the term $n=1$, while the terms corresponding to clustered sources give rise to complicated expressions after this subtraction.  By now, %we leave the question of convergence, but we will have to address it more carefully in a definitive paper. 

Now, we define
\begin{equation}
\Omega_n=\int_{\Omega_b}
\,\ldots \int_{\Omega_b}  \prod_{j=1}^{n} h^{\eta-1}(\mathbf{x_j}) \,w_n(\mathbf{x_1},...,\mathbf{x_n}) \, \mathbf{dx_1} \ldots \mathbf{dx_n}.
\end{equation}
\noindent
 $\Omega_n$ is an integral incorporating the information about the beam pattern and the correlation. 
 Putting together all these integrals, we arrive at the following formula for the Fourier transform 
\begin{equation}  \label{eq:eq7}
\phi(\omega)=\exp \sum_{n=1}^{\infty} \, {-\gamma_n} |\omega|^{n\alpha}(1+i\, \sgn(\omega)\tan(n\pi\alpha /2)),
\end{equation}
\noindent 
with
\begin{equation} \label{eq:eq8}
 \gamma_n=-  \frac{k^n \Gamma^n(-\alpha) \Omega_n \cos(n\pi\alpha/2)}{n!}.
\end{equation}
\noindent
If we consider the special case $n=1$ (no correlations) we recover, after some simple algebra, formula (7) of \cite{Herranz:2004}.
Handling the formula above in a general case is very complicated. In order to simplify matters we will only consider the two-point correlation, i.e. we neglect higher-order correlations from now on (see Subsection 2.2), $ \gamma_n=0$ for $n>2$.
All the formulas derived in this paper are valid for generic beam profiles and generic two-point correlation functions, provided that they are written in terms of $\Omega_1$ and $\Omega_2$. However, for the sake of simplicity, we  will assume that the beam profile is Gaussian, (we will also assume a small beam and therefore work on the plane)
\begin{equation}
 h(\theta)=e^{-4 \log{2}\,(\theta/\theta_b)^2},
\end{equation}
\noindent
with $\theta_b$  the full width half maximum (FWHM) of the beam. With this assumption the calculation of $\Omega_1$ is straightforward
\begin{equation}
\Omega_1= \frac{\pi \theta_b^2}{4\log{2}\,\alpha}.
\end{equation}
\noindent
To compute $\Omega_2$ we need to specify a formula for the two-point angular correlation. We will use a typical power law, see formula (67)  in \cite{Takeuchi} and Section~\ref{sec:counts}
\begin{equation}
w_2(\mathbf{x_1},\mathbf{x_2})=\left( \frac{\theta_{12}}{\theta_0}\right)^{-\delta},
\end{equation}

\noindent
where $\theta_{12}$ is the angular distance between the points and $\theta_0$ the characteristic correlation angular distance. By substituting this formula in $\Omega_2$, we obtain
\begin{equation}
\Omega_2=2\pi \theta_0^\delta \left( \frac{\theta_b^2}{4\log{2}\, \alpha}\right)^{(4-\delta)/2}\Gamma\left(\frac{4-\delta}{2}\right) G(\delta)
\end{equation}
\begin{equation}
G(\delta)=\int_0^{\pi/2}\, \int_0^{\pi/2}\,  \frac{\sin{x}\,dx\,dy }{(1+\sin{x}-2\sin{x}\sin^2{y})^{\delta/2}}.
\end{equation}
\noindent
This last integral converges for $0 < \delta <2$ and can be worked out analytically
\begin{equation}
G(\delta)=\frac{\pi  \, 2^{-\delta/2}}{(2-\delta)}. 
\end{equation}
\noindent
We further simplify the formula for $\Omega_2 $ and find
\begin{equation}
\Omega_2= \frac{\pi^2}{4} \theta_0^\delta \left( \frac{\theta_b^2}{2\log{2}\, \alpha}\right)^{(4-\delta)/2}\Gamma\left(1-\frac{\delta}{2}\right). 
\end{equation}
\noindent
Thus, if we knew the parameters $\alpha, k, \theta_0, \delta$ and $\theta_b$, we could obtain $\phi(\omega)$ and, applying the inverse Fourier tranform, the pdf $f(I)$. $\phi(\omega)$, as given in formula (9) with $\gamma_n=0$ for $n>2$, is the characteristic function of the sum of two independent random variables with alpha-stable distibutions $S_{\alpha}(1,\gamma_1,0)$ 
and $S_{2\alpha}(1,\gamma_2,0)$ respectively. For the characteristic function to be invertible $\gamma_2$ must be positive, i.e.  $1/2 < \alpha <3/2$. $\gamma_1$ is always positive, since $0 <\alpha < 1$ in our calculations.

\vspace{5mm} 
 
\section{Logarithmic estimators} \label{sec:logarithmic}

Our goal is to determine the unknown parameters $\alpha$, $k$, $\theta_0$, and $\delta$ from simulations based on real data and, in order to do that, we will follow the scheme presented in \cite{Herranz:2004}, that is, we will make use of logarithmic estimators. The logarithmic moments of our distribution
can be found by using
\begin{equation}
E\left((\log |I|)^k\right)  = \frac{d^k}{dp^k}\left(E\left(|I|^p\right)\right)_{p=0} .
\end{equation}
\noindent
The right-hand side of this formula can be obtained in the case of an alpha-stable distribution by means of Zolotarev's Theorem \cite{Zolotarev,Kuruoglu01}, that allows us to calculate $E\left(|I|^p\right)$. We generalize Zolotarev's theorem in Appendix~\ref{sec:appendixA} for the sum of two independent random variables with alpha-stable distributions. For our particular case, $\alpha_2=2\alpha_1$, we must compute the following integral 
\begin{equation} \label{eq:diecisiete}
E\left(|I|^p\right)= \frac{p}{\alpha \Gamma(1-p)\cos(p\pi/2) } \, \real \left[  \, \int_{0}^{\infty} \,  \frac{1-e^{-au-bu^2}}{u^{p/\alpha+1}} \, du \right] 
\end{equation}
with $a=\gamma_1(1+i\tan(\alpha\pi/2))$,   $b=\gamma_2(1+i\tan(\alpha\pi))$, the integral converges if $\gamma_1$ and $\gamma_2$ are positive  ($1/2<\alpha < 1$) and for $0 < p/\alpha <1$. In Zolotarev's theorem, $b=0$.
After performing this integral (see Appendix~\ref{sec:appendixA}), we find an extension of Zolotarev's theorem
\begin{equation} \label{eq:eIp1}
E(|I|^p )= \frac{\Gamma(1-p/\alpha)}{\Gamma(1-p)\cos(p\pi/2) } \real \left[\, b^{p/(2\alpha)} H(p/\alpha,z)\right],
\end{equation} 
\noindent
where $H(p/\alpha,z)$ is the Hermite function \cite{Lebedev1972}
and $z= a/2\sqrt{b}$.
We can also express this formula in the following way
\begin{equation}
E\left(|I|^p\right)  =  \frac{p}{\alpha \Gamma(1-p)\cos(p\pi/2) } \real \left[ M\left(\alpha,p \right)  \right],
\end{equation}
\noindent
where
\begin{equation}
M\left(\alpha,p \right) = b^{p/(2\alpha)} \left[ z \Gamma
\left( \frac{1}{2} - \frac{p}{2\alpha} \right) \, _1F_1\left(\frac{1}{2}-\frac{p}{2\alpha};\frac{3}{2};z^2\right)  
 -  \frac{1}{2} \, \Gamma\left(-\frac{p}{2\alpha}\right) \, _1F_1\left(-\frac{p}{2\alpha};\frac{1}{2};z^2\right) \right]
\end{equation}
%where
%\begin{equation}
%\begin{aligned}
%K(p/\alpha,z)=z\Gamma(1/2-p/(2\alpha))_1F_1(1/2-p/(2\alpha);3/2;z^2)-  \\ 
%1/2\Gamma(-p/(2\alpha))_1F_1(-p/(2\alpha);1/2;z^2)
%\end{aligned}
%\end{equation}
\noindent
with $_1F_1$ the confluent hypergeometric function or Kummer function 
\cite{Lebedev1972}. However, we prefer to use the Hermite function since
the notation is more compact and the calculations simpler.
If we perform the limit of (\ref{eq:eIp1}) as $|z|$ tends to $\infty $ (|b| goes to zero), we find
\begin{equation} \label{eq:theozol}
E\left(|I|^p\right) = \frac{\Gamma(1-p/\alpha)}{\Gamma(1-p)\cos(p\pi/2) }\, \real \left[ a^{p/\alpha} \right],
\end{equation} 
and we recover the formula for the case without correlations,  Zolotarev's theorem, \cite{Zolotarev,Kuruoglu01}. Our next step is the differentiation of (\ref{eq:eIp1}) to obtain the expectation and variance of $\log(|I|)$:
\begin{equation} \label{eq:eIp2}
E(\log {|I|})=\left(\frac{1}{\alpha}-1\right) \gamma+\frac{1}{2\alpha}\log{ \frac{k^2\Gamma^2(-\alpha)\Omega_2}{2}}+
\frac{1}{\alpha} \real \left[ \frac{dH}{du}(0,z) \right]
\end{equation}
\begin{eqnarray} \label{eq:varlogI}
\mathrm{var} (\log {|I|}) & =& \frac{\pi^2}{6}\left(1/2+\frac{1}{\alpha^2}\right)-\frac{\theta^2}{4\alpha^2} +
\frac{1}{\alpha^2} \real \left[ \frac{d^2H}{du^2}(0,z)\right] 
- \frac{1}{\alpha^2} \left\lbrace \real \left[\frac{dH}{du}(0,z) \right]\right\rbrace^2
\nonumber \\
& - &  - 
 \frac {\theta}{\alpha^2} \mathrm{Im} \left[ \frac{dH}{du}(0,z) \right].
\end{eqnarray}
\noindent
In these formulas $\gamma=0.5772...$ is the Euler constant, $\theta=\arctan(\tan(\alpha \pi))$,  $u=p/\alpha $ and $z$ can be written as
\begin{equation}
z= \frac{\Omega_1}{\sqrt{2\Omega_2}} \, i=\left( \frac{ \theta_b^2}{ 2 \log{2} \alpha \, \theta_0^2}\right)^{\delta/4} \frac{1}{\sqrt{2\Gamma{(1-\delta/2)}}} \, 
 \, i.
\end{equation}
Equation
(\ref{eq:eIp2}) can be expanded by substituting $\Omega_2$ in terms of $\Omega_1$ and $z$
\begin{equation} \label{eq:eIp3}
E(\log {|I|})=\left(\frac{1}{\alpha}-1\right)\gamma+\frac{1}{\alpha}\log{(k |\Gamma(-\alpha)| \Omega_1)}+
\frac{1}{\alpha} \left\lbrace \real \left[\frac {dH}{du}(0,z)\right]-\log(2|z|)\right\rbrace.
\end{equation}
\noindent
Both (\ref{eq:varlogI}) and (\ref{eq:eIp3})  can be computed as $|z|$ approaches $\infty$, leading us to the formulas corresponding to the distribution of uncorrelated sources (formulas (21) and (22) in \cite{Herranz:2004}).
 If we are interested in calculating the cumulants of $\log{|I|} $ in general, we can use the expression for $ \log {E\left(|I|^p\right)} $ in terms of the Hermite functions and take the successive derivatives at $p=0$. If we define
  \begin{equation}
  r= \frac{k^2\Gamma^2(-\alpha)\Omega_2}{2}
  \end{equation}
  and separate the Hermite function in its real and imaginary part, $H_1+iH_2$, we can write
  \begin{eqnarray}
     \log{E\left(|I|^p\right)} & =& \log{\Gamma(1-p/\alpha)}-\log{\Gamma(1-p)}+\log{\left(\sec(p\pi/2)\right)}+\log{\left(\cos\left( \frac{p\,\theta}{2\alpha}\right)\right)}+  \frac{p}{2\alpha} \log{r} \nonumber \\ & + &  \log{\left(H_1-\tan{\left( \frac{p \, \theta}{2\alpha}\right)}H_2\right)} .
    \end{eqnarray}
\noindent  
  Now, we can differentiate at $p=0$ and obtain the corresponding cumulants
  \begin{equation} \label{eq:lakappa1}
  \kappa_1=\psi(1)\left(1-\frac{1}{\alpha}\right)+ \frac{1}{\alpha} \left( \frac{\log{r}}{2}+{H_1^{\prime}}(0,z)\right) 
  \end{equation}
  This is the mean, the same as (\ref{eq:eIp2}). The following cumulant, the variance, is given by  
  \begin{equation} \label{eq:lakappa2}
    \kappa_2=\psi^{(1)}(1)\left(\frac{1}{\alpha^2} -1\right)+ \frac{\pi^2}{4}+ \frac{1}{\alpha^2} \left({H_1^{\prime \prime}}(0,z)-{H_1^{\prime}}^2(0,z)-\theta {H_2^{\prime}}(0,z)-\frac{\theta^2}{4}\right),
    \end{equation} 
    \noindent
    where $\psi^{(n-1)} (x)$ is the polygamma function, defined as 
\begin{equation}    
    \psi^{(n-1)} (x)= \frac{d^n\, \log\Gamma(x)}{dx^n}, \, \,  \, \psi^{(0)}=\psi, \, \, \, 
    \psi(1)=-\gamma, \, \,  \,  \psi^{(1)}(1)=\pi^2/6. 
\end{equation}
\noindent 
    The third-order cumulant is
    \begin{eqnarray} \label{eq:lakappa3}
        \kappa_3 & =& \psi^{(2)}(1)\left(1-\frac{1}{\alpha^3}\right) \nonumber \\ & + & \frac{H^{\prime \prime \prime}_1(0,z)-3 H_1^{\prime \prime} (0,z) {H_1^{\prime}} (0,z)+2 {H_1^{\prime}}^3(0,z)+3\theta {H_2^{\prime}}(0,z) {H_1^{\prime}}(0,z)- 
        3\theta \frac{H_2^{\prime \prime}(0,z)}{2}}{\alpha^3} . 
        \end{eqnarray}
        \noindent 
        The expression for the fourth-order cumulant is too long to be written here, but quite straightforward to obtain. Besides, we can use the series expansion of the Hermite functions \cite{Lebedev1972}  for calculating the derivatives of $H_1$ and $H_2$.
         \begin{equation} \label{eq:hermite}
                        H(\nu,z)= \frac{\sqrt{\pi} \,2^{\nu}}{\Gamma(1/2-\nu/2)}- \frac{\nu}{2 \Gamma(1-\nu)}\,\sum_{n=1}^{\infty}  \frac{ \Gamma(n/2-\nu/2)(-2z)^{n}}{n!}.
                        \end{equation}
\noindent 
         For instance
        \begin{equation}
        {H_1^{\prime}}(0,z)=\psi(1)/2-\,\sum_{n=1}^{\infty}  \frac{(-1)^n \Gamma(n)(2|z|)^{2n}}{ 2\, (2n)!}
        \end{equation}
    \begin{equation}
            {H_1^{\prime \prime}}(0,z)= \frac{1}{4}(\psi^2(1)-\psi^{(1)}(1/2))- 
            \,\sum_{n=1}^{\infty}  \frac{(-1)^n \Gamma(n)(2|z|)^{2n} (\psi(1)-\psi(n)/2)}{ \, (2n)!}
            \end{equation}
            \begin{equation}
                    {H_2^{\prime}}(0,z)=\sum_{n=0}^{\infty}  \frac{(-1)^n \Gamma(n+1/2)(2|z|)^{2n+1}}{ 2\, (2n+1)!}.
                    \end{equation}
  \noindent
 Higher order derivatives can be computed by iterative differentiation, giving rise to cumbersome formulas.

\vspace{5mm}

\section{Symmetric case } \label{sec:symmetric}

 For the sake of simplicity, we will now consider symmetrized sequences defined by 
 \begin{equation}  \label{eq:symm}
 I_k^s=I_{2k}-I_{2k-1} ,
 \end{equation}
\noindent 
for $k=1,\ldots,N/2$ and $N$ the number of independent data. It can be readily proved that in this case $a=2\gamma_1$ and $b=2\gamma_2$, so that the imaginary part of  formula (9) vanishes. As before, we assume $ \gamma_n=0$ for $n >2$, i.e. we only take into account two-point correlations. Please note that, in order to avoid correlations in the case of real data, we must consider data from distant enough pixels before applying (\ref{eq:symm}). We can obtain the expressions for the logarithmic cumulants by using Hermite functions. Now, since $a$ and $b$ are real, so are 
\begin{equation}
z= \frac{a}{2\sqrt{b}}= \frac{\Omega_1\, \cos(\alpha \pi/2)}{\sqrt{-\Omega_2 \cos(\alpha \pi)}}
\end{equation}
\noindent and the Hermite function. 
This fact greatly simplifies the formulas, making these difference maps very useful for the determination of the parameters.  
If we define 
\begin{equation}
\sigma= \frac{1}{\left( \frac{\theta_0}{\theta_b}\right)^{\delta/2} (2\,\log2\, \alpha)^{\delta/4} \,\sqrt{\Gamma(1-\delta/2)}},
 \end{equation}
\noindent  
 then $\Omega_2=( \Omega_1/\sigma)^2$ and $z= \sigma \cos{(\alpha \pi/2)} /
\sqrt{-\cos{(\alpha \pi)}}$,
and now we have
\begin{equation}  \label{eq:logEIsp}
     \log{E(|I^s|^p)}=\log{\Gamma(1-p/\alpha)}-\log{\Gamma(1-p)}+\log{sec(p\pi/2)}+ \frac{p}{2 \alpha} \log{b} + \log{H(p/\alpha,z)},
    \end{equation}
\noindent    
    with
  \begin{equation}
    b=-k^2\Gamma^2(-\alpha)\Omega_2 \cos{\alpha \pi}.
    \end{equation}  
    Bear in mind that since $ 1/2< \alpha < 1 $, $b$ is positive. Now, we can write the cumulants in a very simple way by taking succesive derivatives of (\ref{eq:logEIsp}) at $p=0$
  \begin{equation} \label{eq:kappa1}
   \kappa_1=\psi(1)\left(1-\frac{1}{\alpha}\right)+ \frac{1}{\alpha} \left( \frac{\log{b}}{2}+{H^{\prime}}(0,z)\right)
   \end{equation}   
\begin{equation} \label{eq:kappa2}
    \kappa_2=\psi^{(1)}(1)\left(\frac{1}{\alpha^2}-1\right)+ \frac{\pi^2}{4}+ \frac{1}{\alpha^2} \left({H^{\prime \prime}}(0,z)-{H^{\prime}}^2(0,z)\right)
    \end{equation}
\begin{equation}     \label{eq:kappa3}
        \kappa_3=\psi^{(2)}(1)\left(1-\frac{1}{\alpha^3}\right)+ \frac{1}{\alpha^3} \left(H^{\prime \prime \prime}(0,z)-3{H^{\prime \prime}}(0,z) {H^{\prime}}(0,z)+2 {H^{\prime}}^3(0,z)\right) 
        \end{equation}
        \begin{eqnarray}  \label{eq:kappa4}
                \kappa_4 & =& \psi^{(3)}(1)\left(\frac{1}{\alpha^4}-1\right)+ \frac{\pi^4}{8} \nonumber \\ & + & \frac{H^{\prime \prime \prime \prime}(0,z)-4H^{\prime \prime \prime}(0,z) {H^{\prime}}(0,z)-3 {H^{\prime \prime}}^2(0,z)+ 
                12{H^{\prime \prime}}(0,z){H^{\prime}}^2(0,z) -6{H^{\prime}}^4(0,z)}{\alpha^4} 
  \end{eqnarray}
\noindent                
                From (\ref{eq:kappa2}) and (\ref{eq:kappa3}) we could obtain $\alpha$ and $\sigma$ ($\theta_0$ and $\delta$ cannot be separated, and we must fix one, typically $\delta$, to determine the other). The variable $k$ can be isolated 
                in (\ref{eq:kappa1}), because it only appears in $\kappa_1$ inside $b$.
                
    It would be possible to write more cumulants, by following the general pattern given by the polygamma functions, the $n$-th derivative of $ \tan(p\pi/2)$ at $p=0$ and taking into account that the part involving the derivatives of $H(p/\alpha,z) $ follows the same pattern as the cumulants, expressed in terms of the derivatives of $H$ instead of the raw moments. The derivatives of $H$ can be calculated from the corresponding series (\ref{eq:hermite}). For instance
    \begin{equation}
            {H^{\prime}}(0,z)=\psi(1)/2-\,\sum_{n=1}^{\infty}  \frac{\Gamma(n/2)(-2z)^{n}}{ 2\, n!}
            \end{equation}
            \begin{eqnarray}              
                        {H^{\prime \prime}}(0,z) & =&  \frac{1}{4}\left(\psi^2(1)-\psi^{(1)}(1/2)\right) \nonumber \\ & - & 
                        \,\sum_{n=1}^{\infty}  \frac{ \Gamma(n/2)(-2z)^n (\psi(1)-\psi(n/2)/2)}{ \, n!} .
                        \end{eqnarray}
    
\vspace{5mm}
    
\section{Simulations and Results} \label{sec:simulations}

%\begin{figure}
%\centering
% \includegraphics[width=0.8\textwidth]{distributions_whittam.pdf}
% \caption{Distribution $f(I)$ for $\alpha=[0.55,0.6,\ldots,0.95]$ and %the parameters $\theta_0=0.06^{\prime \prime}$, $\delta=0.8$, $k=1000$, %FWHM=8$^{\prime \prime}$. The distribution is symmetric around $I=0$; %only the positive part is shown.} \label{fig:fdeI}
%\end{figure}

    From now on, we will try to determine the parameters $ k, \alpha, \sigma $ ( from $\sigma$ we could obtain $\theta_0$ assuming that we know $\delta$ ) by using our formulas for the cumulants and comparing their values with those obtained from simulations based on real data. The pdf of the symmetrized intensity $f(I^s)$ can be calculated as the inverse Fourier transform of $\phi(\omega)$   

    \begin{equation}      
         f(I^s)=  \frac{1}{\pi}  \int_0^{\infty}   \exp{(-a\omega ^{\alpha}-b\omega^{2\alpha})}  \,\cos(\omega I^s) \, d\omega.
         \end{equation}
         
    Please note that this formula is valid for the symmetrized intensity
    $I^s$, but not in the general case, in which we have to deal with the imaginary part. If we integrate this expression, we can calculate the cdf

    \begin{equation}      
         g(I^s)= \frac{1}{2} +  \frac{1}{\pi}  \int_0^{\infty}   \exp{(-a\omega^{\alpha}-b\omega^{2\alpha})}  \,  \frac{\sin(\omega I^s)}{\omega} \, d\omega.
         \end{equation}
\noindent The cumulative                   
$g(I^s)$ can be computed numerically and used for generating simulations. However, we can also use the fact that our pdf is that of the sum of two symmetric alpha-stable variables with distributions $S_{\alpha}(0,a,0)$ and $ S_{2\alpha}(0,b,0)$ respectively. Therefore, we simulate the corresponding alpha-stable variables \cite{Chambers76,Weron} and add them up. We simulate the first variable as
\begin{equation}      
     x_1=a^{1/\alpha} \, \frac{\sin(\alpha t)}{\cos(t)^{1/\alpha}} \, \left[\frac{\cos(t(1-\alpha))}{w}\right]^{(1-\alpha)/\alpha}
     \end{equation}
    \noindent
and the second as
\begin{equation}      
     x_2=b^{1/2\alpha} \, \frac{\sin(2\alpha t)}{\cos(t)^{1/2\alpha}} \, \left[\frac{\cos(t(1-2\alpha))}{w}\right]^{(1-2\alpha)/2\alpha} ,
     \end{equation}
     \noindent
     where $t$ follows a uniform distribution between $-\pi/2$ and $\pi/2$ and w an exponential distribution with mean 1.
We generate these data $x=x_1+x_2$, changing the values of $k$,  $\alpha$, $\theta_0$ and $\delta$ and try to obtain   $k$,  $\alpha$ and $\sigma$ by using the logarithmic cumulants (\ref{eq:kappa1}), (\ref{eq:kappa2}), (\ref{eq:kappa3}). The only caveat is that our results are only valid for $1/2 < \alpha < 1$ 
(however, see \cite{Condon12,Vernstrom,Whittam17a,Whittam17b}).
In the following, we will simulate symmetrized intensities $I_s$  by using the sum of two alpha-stable random numbers, $I_s=x_1+x_2$ , generated according to the counts and correlation parameters chosen from recent observations. We assume that the FWHM of the beam , $\theta_b$, is also known. 

Among recent results on the analysis of the $P(D)$ -- the confusion probability distribution, $P(D)$, of the deflections, $D$ (in Jy/beam), produced by point sources at flux densities below the flux detection limit of the survey -- a partially new approach has been presented by \cite{Vernstrom}, who analyzed a new sample of very faint radio sources. By this approach, ``\textit{with a more robust model, and a comprehensive error analysis}'', \cite{Vernstrom} were able to estimate the $\mu$Jy and sub--$\mu$Jy source counts by using new deep wide-band 3-GHz data in the Lockman Hole from the Karl G. Jansky Very Large Array (VLA). These authors found that the differential source counts between $0.05 \mu$Jy and $0.2 \mu$Jy  present a slope of $\eta=1.79$. At the same time, they demonstrated that the source number counts can be constrained down to $\sim$50 nJy, a factor of $\sim$20 below the rms confusion noise.

In the Table 4 of \cite{Vernstrom} the differential number counts follow a power law with $k=1000$ and $\alpha=0.79$ at the lowest density fluxes probed by their observations. The correlation parameters are $\theta_0=0.06^{\prime \prime}$ and $\delta=0.8$; finally the FWHM of the experiment is $8^{\prime \prime}$, \citep{Vernstrom}. As a first application of our method, we use these numbers in a sort of toy model by applying the logarithmic cumulants in two different ways: 
\begin{enumerate}
\item We carry out 100 simulations with the above parameters. Each simulation consists of $10^6$ symmetrized intensities, generated as the sum of alpha-stable random numbers as explained before. The number is large enough to have representative samples, though we have checked that numbers greater than $10^4$ could equally work. In principle, due to the simulation technique, we do not impose any cut in the simulated fluxes although the minimum fluxes are in the nanoJansky range.

For each simulation, we calculate, from our simulated data, the expectation and variance of $\log|I_s|$ and, assuming that we know the correlation parameters $\delta=0.8$ and $\theta_0=0.06''$, we obtain $\alpha$ from (\ref{eq:kappa2}) and then $k$ from (\ref{eq:kappa1}). When we compare the real values of  $k$ and $\alpha$ with those obtained with our method we see
that the mean of the absolute value of the relative error, taking into account our 100 simulations, is less than $0.1 \%$ for $\alpha$  ($0.7\%$ for $k$) .
If we calculate the expectation and variance of $\log|I_s|$ from our simulations and use instead formulas (\ref{eq:eIp2}) and (\ref{eq:varlogI}) of \cite{Herranz:2004} to determine $\alpha$ and $k$,i. e. we assume no correlations, the error in $\alpha$ is still low, $3\%$, though much greater than with our formulas, but the error in $k$ increases clearly, $22\%$. 
%The results are very stable and do not depend on the initial guess of $\alpha$ that we use for the numerical solution of (\ref{eq:kappa2}).
If we simulate with other values of $\alpha$, $ 0.5 <\alpha <1 $,maintaining the value of the other parameters,the errors are very similar.

\begin{table}
  \begin{center}
    \caption{Mean absolute percentage error (MAPE) for the estimation of the $\alpha$ and $k$ parameters as a function of $\alpha$ for one of the two kinds of astrophysical correlation described in the main text \citep{Blake2002}. As a comparison, in the fourth and fifth columns the same quantities are computed using the estimators for the case of no spatial correlation \citep{Herranz:2004}.}
    \label{tab:table1}
    \begin{tabular}{p{1.5cm}p{1.5cm}p{1.5cm}p{1.5cm}p{1.5cm}} 
    \hline
    \hline
     $\alpha$ & MAPE$_{\alpha}$ & MAPE$_{k}$ & MAPE$_{\alpha,nc}$ & MAPE$_{k,nc}$ \\
    \hline
     \hline
%     \multicolumn{5}{c}{Case from \cite{Vernstrom}} \\
%     \hline
%     0.6 & -- & -- & 0.4 &  3.0 \\ 
% 0.7& 0.08& 0.8& 1.2& 9 \\
% 0.8&  0.07& 0.8& 3& 23 \\
% 0.9&  0.06& 0.9& 11& 57 \\
%     \hline
%     \multicolumn{5}{c}{Case from \cite{Blake2002}, assuming $\theta_0$ and $\delta$ are known} \\
%     \hline
    0.6 &  0.12&  1& 2.2& 15 \\
0.7&  0.2& 1.2& 6& 37 \\
0.8&  0.4& 4& 14& 65 \\
0.9&  0.6& 10& 32& 91 \\
    \hline
    \end{tabular}
  \end{center}
\end{table}

% \begin{table}
%   \begin{center}
%     \caption{Mean absolute percentage error (MAPE) for the estimation of the $\alpha$, $\theta_0$ and $k$ parameters as a function of $\alpha$ for two typical cases of spatial correlation. As a comparison, in the fifth and sixth columns $\alpha$ and $k$ are estimated using the estimators for the case of no spatial correlation \cite{Herranz:2004}. In this approximation, it makes no sense to try to estimate $\theta_0$.}
%     \label{tab:table2}
%     \begin{tabular}{p{2cm}p{2cm}p{2cm}p{2cm}p{2cm}p{2cm}} 
%     \hline
%     \hline
%     $\alpha$ & MAPE$_{\alpha}$ & MAPE$_{k}$ & MAPE$_{\theta_0}$ & MAPE$_{\alpha,nc}$ & MAPE$_{k,nc}$ \\
%     \hline
%     \hline
%     \multicolumn{6}{c}{Case from \cite{Vernstrom}, when $\theta_0$ is known only as a order of magnitude} \\
%     \hline
% 0.7 &  0.2& 2&14&  1.2& 9\\
% 0.8&  0.3& 3& 17&  3& 23\\
% 0.9&  0.2& 9& 32&  11& 57 \\
%     \hline
%     \multicolumn{6}{c}{Case from \cite{Blake2002}, when $\theta_0$ is known only as a order of magnitude} \\
%     \hline
% 0.6&  0.2& 1.8& 7&  2.2& 15\\ 
% 0.7&  0.2& 2& 8&  6& 37 \\
% 0.8&  1.8& 14& 46&  14& 65 \\
% \hline
%     \end{tabular}
%   \end{center}
% \end{table}

In this first case (knowledge of the correlation), we have chosen a second example, increasing the correlation distance $\theta_0=0.64^{\prime \prime}$ as in \cite{Blake2002} and keeping the other parameters. Now the errors in $\alpha$ and $k$ are
$0.4 \%$ and $4\%$ respectively, for $\alpha=0.79$, $k=1000$, whereas they are $14\%$ and $65\%$ if we do not consider the correlations and apply the formulas in \cite{Herranz:2004} to estimate the parameters. 

We also calculate the relative errors for different values of $\alpha$ with the \cite{Blake2002} correlation parameters.  
Table~\ref{tab:table1} shows that the errors are much higher if we do not take into account the real correlation and use the formulas in \cite{Herranz:2004}. 
% The method works well for the determination of $\alpha$ and $k$ even if the angular scale of correlation is known only approximately, as it is shown in Table~\ref{tab:table2}. For this case, even if the estimation of $\theta_0$ is relatively poor, the mean errors in $\alpha$ and $k$ are still small, and much smaller than if we ignore the effect of correlation.

Finally, in Figures~\ref{fig:k1} and~\ref{fig:k2} we plot $\kappa_1$ and $\kappa_2$ obtained from (\ref{eq:kappa1}) and (\ref{eq:kappa2}), for the \cite{Vernstrom} and \cite{Blake2002} cases, with values of $\alpha$ ranging from $0.5$ to $1$ and using the cited values for the other parameters. We can see that these cumulants differ more from the no correlation case the greater $\alpha$ and the correlation distance are. For $\alpha=0.5$, we recover the no correlation result, since $\gamma_2=0$. These figures are plotted just by substituting  the parameters in the formulas, without the use of simulations.

In these examples, we have performed the simulations and recovered $\alpha$ and $k$ assuming a perfect knowledge of the correlations, which will not be, in general, the real case. However, we want to emphasize with our toy model that we could  make large errors in the determination of these parameters if we do not consider the correlations at all, as can be seen in Table 1.

\begin{figure}
\centering
 \includegraphics[width=0.8\textwidth]{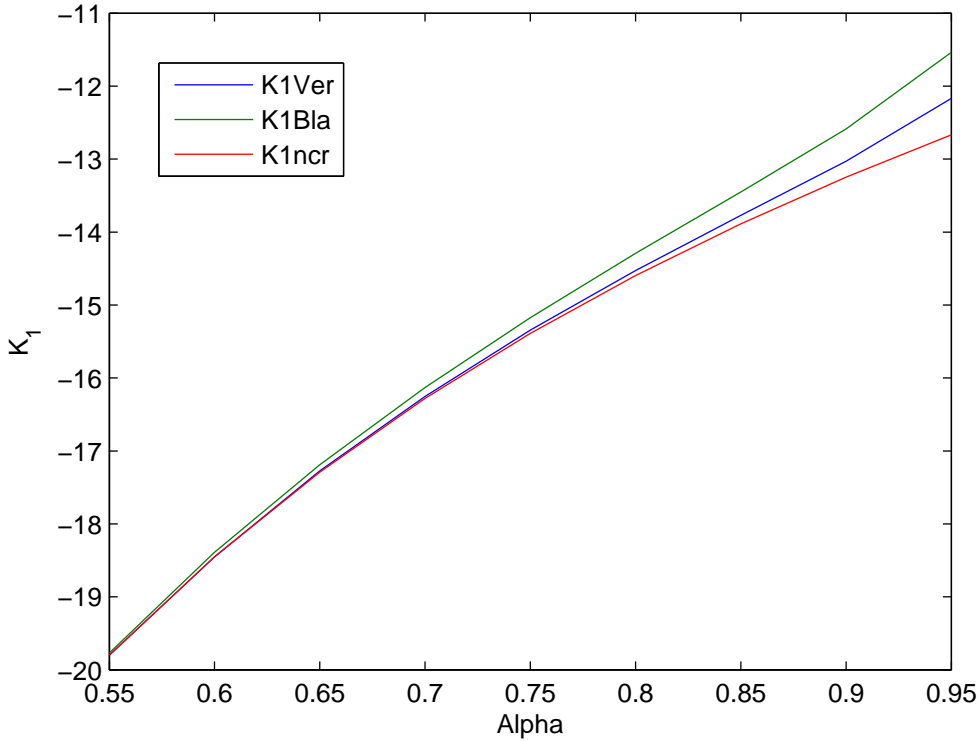}
 \caption{Cumulant $\kappa_1$, as obtained from (\ref{eq:kappa1}), as a function of $\alpha$ for three typical cases: the correlation function from Vernstrom et al. (2014) (blue line), the correlation function from Blake and Wall (2002) (green line) and a case with no spatial correlation (red line). } \label{fig:k1}
\end{figure}

\begin{figure}
\centering
 \includegraphics[width=0.8\textwidth]{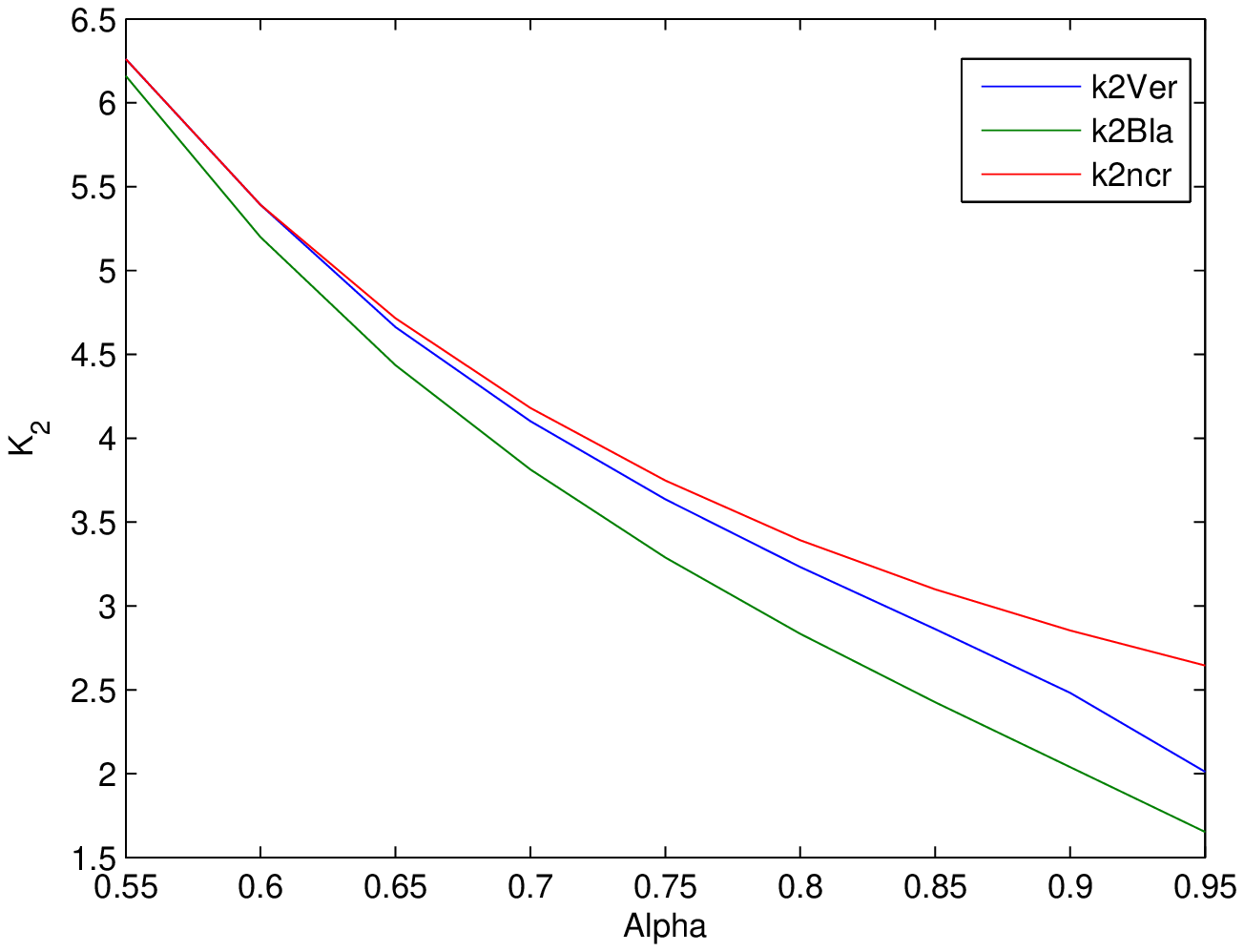}
 \caption{Cumulant $\kappa_2$, as obtained from (\ref{eq:kappa2}), as a function of $\alpha$ for three typical cases: the correlation function from Vernstrom et al. (2014) (blue line), the correlation function from Blake and Wall (2002) (green line) and a case with no spatial correlation (red line). } \label{fig:k2}
\end{figure}

\item Now, we will assume, more realistically,  that we know perfectly $\delta$ but we do not know  $\theta_0$. As in case 1, we carry out 100 simulations with the \cite{Vernstrom} parameters, each  consisting of $10^6$ symmetrized intensities and now, we calculate the cumulants for each simulation, solve (\ref{eq:kappa3}) and (\ref{eq:kappa2}) to obtain $\theta_0$ and $\alpha$, assuming $\delta=0.8$,  and, as before, apply (\ref{eq:kappa1}) to compute $k$. The mean absolute percentage errors  in $\alpha$ and $k$ are $0.3\%$ and $2\%$ respectively, but the error in $\theta_0$ is greater, $13\%$. 

 For the \cite{Blake2002} correlation, we simulate with $\delta=0.8$ and $\theta_0=0.64''$. We proceed as before, calculating the logarithmic cumulants from the simulations and solving (\ref{eq:kappa3}), (\ref{eq:kappa2}) and 
(\ref{eq:kappa1}) to determine $\alpha$, $k$ and $\theta_0$, assuming $\delta=0.8$. The errors are now $1.6\%$, $10\%$ and $43\%$ for $\alpha$, $k$ and $\theta_0$.

These examples show that, in general, a certain knowledge of the correlation, even if we only know its slope, enables us to obtain a much better determination of the two parameters $k$ and $\alpha$, though in some cases the determination of the correlation characteristic angle is not very precise.

For the application of this method to real data, we should symmetrize the data, calculate the logarithmic cumulants from the symmetrized data and solve the system consisting of (\ref{eq:kappa1}), (\ref{eq:kappa2}) and 
(\ref{eq:kappa3}), assuming a certain value of $\delta$.

An important issue is the influence of the instrumental noise in our results. It could be incorporated in the calculation of the characteristic function, formula (\ref{eq:eq7}), just multiplying by the characteristic function of the noise, e.g. in the case of Gaussian noise $\phi_g(\omega)=e^{-\sigma_g^2 \omega^2/2}$. However, the calculation of the logarithmic cumulants in this case would be quite complicated. Logarithmic cumulants and minimum distance estimators \citep{Ilow98}  were used in \cite{Herranz:2004} to study the influence of Gaussian noise in their results. Their main conclusion was that if the noise dispersion $\sigma_g$ is low enough, the logarithmic cumulants, calculated without taking the noise into account, can be safely used to determine the parameters defining the counts. Otherwise, minimum distance estimators, that try to fit the characteristic function, perform better.
In our paper, we intend to present the first method that allows the computation
of the logarithmic cumulants in the clustering case and their use to estimate the counts and correlation parameters. We leave a careful study 
of the determination of the parameters in the clustering plus noise case for a further paper.

\end{enumerate}

\vspace{5mm}

\section{Conclusions} \label{sec:conclusions}

\begin{enumerate}
\item We study the characteristic function --the Fourier transform of the probability density function (pdf)-- of the intensity produced by clustered extragalactic point sources filtered with a typical observational beam. We follow previous works by \cite{Barcons92} and \cite{Takeuchi} and find a new general expression for this characteristic function (\ref{eq:eq7}), (\ref{eq:eq8}), in the case of a power-law distribution for the differential number counts.

\item We analyze carefully the two-point correlation case $ \gamma_n=0 $ for $n> 2$ in (\ref{eq:eq7}). Then, the observed intensity can be written as the sum of two independent random variables that follow alpha-stable distributions $ S_{\alpha}(\gamma_1,1,0)$, $ S_{2\alpha}(\gamma_2,1,0)$. We generalize Zolotarev's theorem (\ref{eq:eIp1}) and prove this generalization in Appendix~\ref{sec:appendixA}. Finally, we write formulas for the cumulants of the logarithm of the absolute value of the intensity (\ref{eq:lakappa1}), (\ref{eq:lakappa2}), (\ref{eq:lakappa3}).

\item We pay special attention to the symmetric case, where we subtract intensities at different points. In this situation, the general logarithmic cumulants happen to be quite simple (\ref{eq:kappa1})-(\ref{eq:kappa4}). These formulas lead us to a general method for determining the parameters defining the differential number counts and the correlation.

\item Finally, we simulate intensity data with power law differential number counts and typical two-point correlation functions and show that the logarithmic cumulants enable us to calculate the parameters $k$, $\alpha$ and $\theta_0$ with small errors, assuming that we know $\delta$. We use the counts and correlations of \cite{Vernstrom} and \cite{Blake2002}, giving the relative errors in the text and in Table~\ref{tab:table1}. In general, taking into account the correlation in the use of the logarithmic cumulants allows a better determination of the basic parameters.
\item The method can be applied to real data in a straightforward way. The data must be symmetrized, then the logarithmic cumulants can be calculated from the data and the parameters defining the counts $k$, $\alpha$ and the correlation angle $\theta_0$ found by solving (\ref{eq:kappa1}), (\ref{eq:kappa2}) and 
(\ref{eq:kappa3}).   

\end{enumerate}

As a final remark, we can say that we have defined a new method to determine essential parameters related to the distribution of clustered astrophysical sources based on the calculation of logarithmic cumulants.
The important issue of how to apply or extend the method to take into account the observational noise deserves a future detailed analysis.

\vspace{5mm}

\section*{Acknowledgements}

We thank the Spanish MINECO
for financial support under projects AYA2015-64508-P and AYA2015-65887-P. 
D. Herranz also thanks 
funding from the European Union's Horizon 2020 research and innovation
programme (COMPET-05-2015) under grant agreement number
687312 (RADIOFOREGROUNDS). Finally, we thank an anonymous referee for his/her very useful suggestions.

% The best way to enter references is to use BibTeX:

\bibliographystyle{aasjournal}
\bibliography{confusion_noise} % if your bibtex file is called example.bib

\newpage 

%%%%%%%%%%%%%%%%%%%%%%%%%%%%%%%%%%%%%%%%%%%%%%%%%%

%%%%%%%%%%%%%%%%% APPENDICES %%%%%%%%%%%%%%%%%%%%%

\appendix

\section{AN EXTENSION OF ZOLOTAREV'S THEOREM.} \label{sec:appendixA}

\bigskip

\subsection*{An extension of Zolotarev's theorem}

If $X_1$ and $X_2$ are random independent variables that follow alpha-stable distributions   $S_{\alpha_1}(\beta_1,\gamma_1,0)$,
$S_{\alpha_2}(\beta_2,\gamma_2,0)$, respectively, with $k=\alpha_2/\alpha_1>1$ and both $\alpha_1$, $\alpha_2 \neq 1 $, then if $X=X_1+X_2$ 
\begin{equation} \label{eq:A1}
E(|X|^p )=\frac{\Gamma\left(1-p/\alpha_1\right)}{\Gamma\left(1-p\right)\cos(p \, \pi/2) } \, \mathrm{Real} \left[ b^{p/(k\alpha_1)} G_k\left(p/\alpha_1,z\right)\right],
\end{equation} 
with $0<p<\alpha_1$, $a=\gamma_1(1+i\beta_1 \tan{(\alpha_1\pi/2)} )$, $b=\gamma_2(1+i\beta_2 \tan{(\alpha_2\pi/2)} )$, $z=a/(2b^{1/k})$ and
\begin{equation}
G_k(\nu,z)=\frac{\Gamma(1-\nu/k)}{\Gamma(1-\nu)}-
\frac{\nu}{k \, \Gamma(1-\nu)} \, \sum_{n=1}^{\infty} \frac{ \Gamma\left(n/k-\nu/k\right) \, (-2z)^{n}}{n!}
\end{equation}
For $k=2$, $G_k(\nu,z)=H(\nu,z)$, the Hermite function and we recover (\ref{eq:eIp1}). If $p=1$ we must substitute $\pi/2$ for $\Gamma(1-p)\cos(p\pi/2) $ in (\ref{eq:A1}). Finally, (\ref{eq:A1}) is also valid if $\alpha_1=1\, (\alpha_2=1) $ provided $\beta_1=0$ $(\beta_2=0)$.

\medskip

\begin{proof}

First we calculate
\begin{equation}
\int_0^{\infty}  \frac{1-\cos{(Xt)}}{t^{p+1}} \, dt
\end{equation}
with the change$|X|t=u$, we obtain
\begin{equation}
\int_0^{\infty} { \frac{1-\cos{(Xt)}}{t^{p+1}} \, dt}=|X|^p \displaystyle\frac{\Gamma(1-p)\, \cos{(p\pi/2)}}{p}
\end{equation}
This result is valid for $0<p<2,\, p\neq 1$, however for $p=1$ we must substitute $\pi/2$ for $\Gamma(1-p)\, \cos(p\pi/2)$ in the calculations.
We assume $p \neq 1$ from now on. Then
\begin{equation} \label{eq:A5}
E\left(|X|^p\right)=\frac{p} {\Gamma(1-p)\, \cos(p\pi/2)}  \int_0^{\infty} { \displaystyle\frac{E(1-\cos{(Xt)})}{t^{p+1}} \, dt}
\end{equation}
Taking into account that $E(\cos(Xt))=E(\real(e^{-iXt}))$
we rewrite (\ref{eq:A5}) as
\begin{equation}
E\left(|X|^p\right)=\frac{p} {\Gamma(1-p)\, \cos(p\pi/2)} \real \left[\int_0^{\infty} { \frac{1-E(e^{-iXt})}{t^{p+1}} \, dt}\right]
\end{equation}
Since $E(e^{-iXt})$ is the characteristic function of $X$ and this is the product of the characteristic functions of $X_1$ and $X_2$, we obtain 
\begin{equation}
E\left(|X|^p\right)=\frac{p} {\Gamma(1-p)\, \cos(p\pi/2)} \real \left[\int_0^{\infty} { \displaystyle\frac{1-e^{-at^{\alpha_{1}}-bt^{\alpha_{2}}}}{t^{p+1}} \, dt}\right]
\end{equation}
With the change $u=t^{\alpha_1}$, we have
\begin{equation} \label{eq:A8}
E\left(|X|^p\right)=\frac{p}{\alpha_1 \Gamma(1-p)\, \cos(p\pi/2)} \real \left[\int_0^{\infty} { \displaystyle\frac{1-e^{-au-bu^k}}{u^{p/\alpha_1+1}} \, du}\right]
\end{equation}
The integral converges for $0< p< \alpha_1$. In the case $k=2$, this is equation (\ref{eq:diecisiete}) of our paper. Finally, we make another substitution, 
$x=b^{1/k}u$. The integral, whose real part we have to calculate, can then be written as
\begin{equation}
I=b^{p/(k\alpha_1)} \,\int_0^{\infty}{ \frac{1-e^{-x^k-2zx}}{x^{p/\alpha_1+1}} \, dx}
\end{equation}
with $z=a \, b^{-1/k}/2$. After this last change, we have to add another integral in the complex plane, but its value is zero, since we integrate along an arc at infinite distance from the origin and the integrand approaches zero in this case.
Our last integral can be readily worked out by expanding $e^{-2zx}$ as a power series. Then we obtain
\begin{equation}
\int_0^{\infty} {\frac{1-e^{-x^k-2zx}}{x^{p/\alpha_1+1}} \, dx}=\int_0^{\infty} { \frac{1-e^{-x^k}}{x^{p/\alpha_1+1}} \, dx}-\sum_{n=1}^{\infty}  \, \frac{(-2z)^n}{n!}\, I_2
\end{equation}
with
\begin{equation}
I_2=\int_0^{\infty} { \frac{e^{-x^k} x^n}{x^{p/\alpha_1+1}} \, dx}=\frac{1}{k} \,
\Gamma\left(\frac{n}{k}-\frac{p}{\alpha_1 k}\right)
\end{equation}
\noindent 
and, integrating by parts 
\begin{equation}
\int_0^{\infty} { \frac{1-e^{-x^k}}{x^{p/\alpha_1+1}} \, dx}=\displaystyle\frac{\alpha_1 }{p} \, \Gamma\left(1 - \frac{p}{\alpha_1 k} \right).
\end{equation}
Puting all the terms together
\begin{equation}
I=b^{\, p/(k\alpha_1)}\,\left[ \frac{\alpha_1 \Gamma(1-p/(k\alpha_1)) }{p}-\sum_{n=1}^{\infty}  \, \frac{(-2z)^n \Gamma\left(n/k-p/(\alpha_1 k)\right) }{ k\, n!}\right]
\end{equation}
 multiplying by $\displaystyle\frac{p}{\Gamma(1-p) \cos{(p\pi/2)}\alpha_1}$, and taking the real part, we reach the final result
\begin{equation}
E\left(|X|^p \right)=\frac{\Gamma(1-p/\alpha_1)}{\Gamma(1-p)\cos(p\pi/2) } \real \left[ b^{p/(k\alpha_1)} G_k(p/\alpha_1,z)\right].
\end{equation}
For $k=2$, we recover the Hermite function (\ref{eq:hermite}), taking into account that
\begin{equation}
\displaystyle\frac{\Gamma(1-p/(2\alpha_1))}{\Gamma(1-p/\alpha_1)}=\displaystyle\frac{\sqrt{\pi} \,2^{p/\alpha_1}}{\Gamma(1/2-p/(2\alpha_1))}. 
\end{equation}
\noindent
If we only have an alpha-stable distribution $X=X_1$, then $b=0$ in (\ref{eq:A8}), this integral can be readily calculated and we find (\ref{eq:theozol}), i.e. Zolotarev's theorem .
As a final note, the theorem can be extended to general alpha-stable distributions $S_{\alpha_1}(\gamma_1,\beta_1,\mu_1)$, $S_{\alpha_2}(\gamma_2,\beta_2,\mu_2)$ if we write $E(|X-\mu_1-\mu_2|^p)$ instead of $E(|X|^p)$ in (\ref{eq:A1}).
\end{proof}

%%%%%%%%%%%%%%%%%%%%%%%%%%%%%%%%%%%%%%%%%%%%%%%%%%

% Don't change these lines
	% typesetting comment
\label{lastpage}
\end{document}